\documentclass[epj]{svjour}

\usepackage{latexsym}
\usepackage{graphics}
\def\eq#1{{Eq.~(\ref{#1})}}

\def\di{\mbox{d}}
\def\ltap{\ \raisebox{-.4ex}{\rlap{$\sim$}} \raisebox{.4ex}{$<$}\ }
\def\gtap{\ \raisebox{-.4ex}{\rlap{$\sim$}} \raisebox{.4ex}{$>$}\ }
\newcommand{\be}{\begin{equation}}
\newcommand{\ee}{\end{equation}}
\newcommand{\bea}{\begin{eqnarray}}
\newcommand{\eea}{\end{eqnarray}}
\newcommand{\nn}{\nonumber}
\begin{document}
\title{Gravitational Interaction of Neutrinos\\
in Models with Large Extra Dimensions}
\author{M.\ Fabbrichesi, M.\ Piai, and G.\ Tasinato}

\institute{INFN, Sezione di Trieste and\\
Scuola Internazionale Superiore di Studi Avanzati\\
via Beirut 4, I-34014 Trieste, Italy.}
\abstract{Whenever fields  are allowed to propagate in different portions
of space-time, the four dimensional theory exhibits
 an effective violation of the principle of equivalence.
We discuss  the conditions under which such
an effect is relevant for neutrino physics.
In the simplest case of  compactification on a flat manifold,
the effect of gravity is many orders of magnitude
too weak and plays no role for solar neutrino oscillation.  Instead, it
 could be important in the study of ultra high-energy neutrinos in cosmic rays.
Gravity could also be relevant for lower energy neutrino processes involving bulk sterile states, if the mechanism of compactification is more subtle than that on torii.
\PACS{
     {04.50.+h}{Gravity in more than four dimensions}   \and
     {14.60.St}{Non-standard-model neutrinos}
    } 
} 
\authorrunning{M.\ Fabbrichesi et al.}
\titlerunning{Gravitational interaction on neutrinos in models with large extra dimensions}
\maketitle
%
%
\section{Introduction}
\label{intro}
Neutrinos are neutrinos because they only interact by means of
weak interactions and gravity. The latter interaction is usually negligible because of
 its intrinsic weakness and the
universal way it acts on all neutrino species.

Attempts to bring gravity into play in neutrino physics have necessarily
assumed some form of explicit violation of the equivalence
principle~\cite{gasperini}. The violation is introduced
by hand within a somehow vague theoretical
framework; the absence of an action
from which to derive the equations of motion
may lead, among other problems,
to  energy non-conservation.

Quite independently of these efforts, in a parallel development,
models with space dimensions in addition to the usual three---and
large enough to be observable
even after compactification---have been
suggested~\cite{extra} as a possible solution to the problem of
the large scale difference between gravity and the standard model.
In these models the standard model  lives in
four-dimensional space time~\cite{RS} and---along the lines of
brane-world models~\cite{HW}---only gravity inhabits the extra
dimensions. The
strength of gravity is re-scaled up to a value not far from
that of the Fermi constant,
and the experimental
weakness of the two body potential of gravity
is  explained by the
space volume of the large extra dimensions, for which the
gravitational coupling at large distances must be divided.

In addition to gravity
also matter can be allowed to propagate in the
extra dimensions. In particular, bulk (sterile)
neutrinos have been suggested and their interaction with the
standard-model neutrinos discussed~\cite{bulknu,DS}. In these models,
bulk neutrinos are coupled to the standard-model neutrinos to give the
latter mass and the the ensuing mixing was argued to be consistent with solar, atmospheric and
other neutrino experimental data involving sterile fermions, before the advent
of  SNO results~\cite{SNO}.

However, the interplay between matter and gravity in the extra
dimensions had not been considered. Since gravity is brought down
to a much smaller energy scale, it is natural to ask  whether its
effect on neutrino physics may become as important as that of
other forces. In this letter we show that  for certain choices of
energy and compactification scales, the effect can be sizable.
This effect is not a mere common renormalization because
particles confined within four dimensions and those allowed to
propagate in the extra dimensions feel a different gravitational
interaction. As we shall see, even though no violation of the
equivalence principle is assumed in the fundamental theory (and
therefore there is no problem with energy conservation),  the
different shapes of the wave functions in the extra dimensions
produce an effective (as opposed to an explicit) violation in the
four-dimensional theory.  This means that the model provides a
consistent framework to discuss effective violations of the
equivalence principle in neutrino physics along the
phenomenological lines of~\cite{gasperini}.

\section{Sterile neutrinos in the bulk}\label{sterili}

Let us consider space-time to consist
of the usual four dimensional Minkoski space plus
$\delta$ space-like extra dimensions, describing a
$\delta$-dimensional compact manifold. For instance, the simplest
possibility is that the compact manifold is a torus,
with all the extra dimensions describing a circle
with radius $R$, small enough to escape experimental observation.

We only consider one of the extra (compact) dimensions to be
large enough to have observable effects, while
the others are at much smaller scales
---as we shall see, for more than one large extra dimension the power law
dependence   of gravitational effects on neutrinos render them
irrelevant---and take its size to be
$R \sim 100 \;\mu$m (that is, $1/R \sim 2 \times 10^{-3} $eV) which is
relevant for the experimental tests of sub-millimeter gravity and satisfies
current upper  bounds~\cite{exp}.

Most of our discussion will be, for the sake of definiteness, about a model  \cite{DS} in which all standard-model
 fields are assumed to be localized
within 4-dimensional space-time (for instance on a 3-brane), while
a fermionic massless standard-model singlet is allowed to propagate in the bulk
with gravity.
The Yukawa coupling of left-handed neutrinos with the
sterile neutrino provides a Dirac mass term $m^{(5)}$
through the Higgs mechanism.
This mass must be  tuned in such a way that the coupling neutrino-bulk fermion
is in the correct range in order to reproduce the neutrino oscillation
phenomenology~\cite{DS}.

Once the Kaluza-Klein (KK)
expansion has been performed, we are left with a tower of
sterile neutrinos in the four-dimensional theory---with masses given by
an integer multiplied by $1/R$ ---coupled to the
standard neutrino via the Dirac mass term $m_D$.
 The phenomenological analysis performed in  \cite{DS} shows that this
model, for suitable values of the four dimensional Dirac mass
term $M_D$, lies in the correct range to reproduce the
Mikheyev-Smirnov-Wolfenstein (MSW) Small Mixing Angle (SMA)
solution to the solar-neutrino deficit, thus providing an elegant
solution to the hierarchy problem.

 The SMA-MSW solution is not compatible with the recent data from SNO
collaboration~\cite{SNO}, and also the possibility that subleading oscillations
involving sterile neutrinos take place seems disfavored from the combined
analisis of all neutrino experiments~\cite{rulingout}. Nevertheless, the model
in~\cite{DS} is still of interest for  its minimality and simplicity. For these reasons, we will use it as a toy-model for illustrative porposes,
since it allows a direct  comparison
between the magnitude of the gravitational and weak intereaction effects.

\section{Matter effects: weak vs. gravitational}
\label{matter}

Weak interactions of neutrinos with matter are not universal:
charged leptons in ordinary matter have all the same flavor (that of the electron),
and therefore charged current interactions of neutrinos
going through a medium like the sun or the earth distinguish the first family
neutrinos $\nu^e_L$  from the other species.
Consequently, the Dirac equation of electron neutrinos propagating
in the medium contains a potential (energy) term of the
\be
V_m = \sqrt{2}\, G_{F}\, \xi_{e} \label{m} \, ,
\ee
with $\xi_e$ electrons per unit volume.

The oscillation phenomenon is then determined not
only by the mass-squared differences and vacuum mixing angles
between the flavors, but also by the density of the medium,
which modifies the effective mixing angles between $\nu^e_L$ and the
other species. In the two neutrino flavor oscillation approximation, the mixing angle in matter $\theta_m$ is given by
\be
\label{msw}
\tan 2\, \theta_m =
\frac {\sin 2\, \theta_0} {\cos 2\, \theta_0
- 2\,p \,V_m\, (\Delta m)^{-2}}\,,
\ee
where $p$ is the energy of the neutrinos (ultra-relativistic
approximation), while $\Delta m^2$ and $\theta_0$ are
the oscillation parameters in vacuum.
The minus sign in the
denominator of~\eq{msw} allows for the possibility
of a resonant transition, leading to the  MSW effect~\cite{MSW}.

On the other hand, neutral current interactions do not affect
the oscillation  because of their universality
(in flavor space): a possible contribution
$V_m^{\prime}$ to the Hamiltonian of the system is there,
but it can be factorized out from the Dirac equation.
This flavor universality  is shared also by
ordinary gravitational interactions thus making its detection impossible.

Several proposals have been put forward in the past  to violate
flavor universality by means of a small violation of the equivalence principle~\cite{gasperini}. This fact is  achieved by  the addition by hand of flavor violating couplings
to gravity, which give rise to a gravitational potential term $V_G$
in the evolution equations, and thus modifying the
mixing angles in matter.

All these proposals come with some unwelcome feature like massive gravitons or energy non-conservation because of the explicit violation of the principle of equivalence.  For this reason, it would be desirable
 to explore
models in which the equivalence principle is not violated explicitly in
the action even though gravity couples differently to different kinds
of neutrinos.
This happens in some models based on the existence of
 more than four space-time dimensions. In this case, the principle of
equivalence applies to the full $D$ dimensional system,
but while matter and other standard model fields are localized
on a four-dimensional manifold,  new  degrees
of freedom can propagate in the whole space, and therefore
their gravitational coupling to matter can
be different from those of standard neutrinos.
>From the four dimensional effective theory point of view,
this shows up as an explicit violation of the
universality of gravitational interaction: the couplings  to gravitons
are different for  standard model fields
and the KK modes of the new, higher dimensional degrees of freedom.

 Gravitational effects of this type are  potentially at work in all
 models where fermions are introduced in
the bulk to reproduce the phenomenology
of the oscillation between SM and sterile neutrinos.
Moreover, in these models, the strenght of the gravitational coupling,
that rules this effect, is not a priori negligible because higher
dimensional effects can render it larger.

In order to show explicitly how this arises, and quantify its
impact on model building, we must  compute the forward scattering
amplitude between matter and neutrinos which lead to~\eq{m},
substituting the exchange of a $W$ boson with
a graviton.
Let us consider as an illustrative example the case $D=5$ with a flat background, with one extra-dimension compactified on a circle of radius $R$.  In this case, the interaction is given by
\be\label{int}
\sqrt{4\, \pi^2\, G^{(5)}}\, h_{\mu\nu}\, T^{\mu\nu} \, ,
\ee
where the 5-dimensional metric tensor is defined as
$g_{\mu\nu} = \eta_{\mu\nu} + 2\, \sqrt{4\, \pi^2\, G^{(5)}} h_{\mu\nu}$, while
$T^{\mu\nu}$ is the Dirac field energy-momentum tensor in flat
space-time, and
\be
 G^{(5)}\equiv1/(2\, \pi\, M_5^3)
\ee
replaces $G_N$ (Newton constant) thus changing the strength of gravitational
interactions.
We consider an idealized situation in which gravity behaves as a five dimensional theory up to a distance equal to $\rho \le R$, and returns back to the usual four dimensional, very weak interaction for distances grater than $\rho$.

The final integration over the
4-dimensional space, having inserted the
correct form of the graviton two-point correlator, gives
\bea
\int \frac{\di^3 x^{\prime}\, \di y^{\prime}\, \delta(y^{\prime})}
{|x^{\prime}-x|^2 + |y^{\prime}-y|^2 }  =
\int\limits_{ {\rm 3-brane}} \frac{\di^3 x^{\prime}}{|x^{\prime}-x|^2
+ y^2} \nn \\
 =   4\, \pi
\label{pot}
\left\{ \sqrt{\rho^2 - y^2} - y \arctan{\sqrt{\rho^2 - y^2}/y} \right\} \, .
\eea
The $\delta$-function in the extra
dimension $y'$ is there because ordinary matter (which we take as
source of the gravitational field) is constrained
within the 3-dimensional space.
We use  the approximation of letting the
gravitational potential act only up to distances of order $\rho$, since  the contribution coming from an integration on larger distances  is negligible.

Finally:
\bea
V_G (y)& =&  - 8\, \pi^{2}\, G^{(5)}\, p \; \xi_N \, m_N \times
\nn\\
&\times&
\left\{ \sqrt{\rho^2 - y^2} - y \arctan{\sqrt{\rho^2 - y^2}/y} \right\}
\label{G}
\eea
The mass $m_N$ is
the rest energy
of the nucleons (of about 1 GeV), while $\xi_{N}$ is their number density
in the medium. Notice in \eq{G}
the extra energy dependence with respect to \eq{m}.

The $y$-dependence of the potential is crucial. For
a system of two neutrinos (like $\nu_e$ and $\nu_\mu$) ,
both constrained inside our 3-dimensional space, the potential  $V_G$
 is only felt at $y=0$, thus giving a common factor
\be
\int V_G(y)\, \delta (y)\, \di y = V_G(0)
\ee
 that can be rotated away in the evolution equation: the principle of
equivalence is here at work.  On the
other hand, bulk neutrinos propagate in the extra dimensions
and feel the whole
potential; accordingly their gravitational interaction is different from
that of ordinary neutrinos.

The Dirac equations for the standard $\nu^e_L$ and bulk
neutrinos, arbitrarily denoted as $N_{L,R}$,
 in five dimensions (and with the $\gamma$ matrices in
chiral representation) propagating in matter of constant density
become:
\be \label{systemeq}
\left\{ \begin{array}{rcl}
\Bigl[ \partial_{0}+ \sigma^{i} \partial_{i}+i\, V_{m} +i\, V_G (y)
\Bigr] \nu_L & =
& - i\, m^{(5)}  N_R \\[0.5em]
\Bigl[ \partial_{0}+ \sigma^{i}{\partial_{i}+i\, V_{G} (y)}
\Bigr]   N_L & = & -
\partial_{y} N_R\\[0.5em]
\Bigl[ \partial_{0}+ \sigma^{i}{\partial_{i}+i\, V_{G} (y)}
\Bigr]   N_R & = &
\partial_{y} N_L + i\, m^{(5)} \nu_L \, ,
\end{array} \right.  \label{dirac}
\ee
where $y$ represents the 4-th space-like coordinate.

\section{Neutrino oscillations}
\label{nuosc}
The evolution equations for the neutrino system, containing the new contributions from gravitational interactions, lead to a modified expression for the mixing angles between standard model neutrino and the tower of KK modes of the five dimensional field.
Let  us  expand the five dimensional field $N_{L}$ in a Fourier series on the fifth coordinate. The field $N_{R}$ decouples from the system and we will not  consider it any more.
The evolution equations of the system can be studied in terms
of the  KK modes,  and we can write it as
\be
i \frac{\di}{\di t} \left(
\begin{array}{c} \nu_L \\ N_L^{(1)} \\ N_L^{(-1)} \\ \cdots \\
 N_L^{(n)} \\ N_L^{(-n)} \\
\end{array} \right) =  {\cal H} \left(
\begin{array}{c} \nu_L \\  N_L^{(1)} \\  N_L^{(-1)} \\ \cdots \\
 N_L^{(n)} \\ N_L^{(-n)}
\end{array} \right) \, , \label{ev}
\ee
where $ {\cal H}$ is defined in \eq{M}.

\begin{figure*}[h,t]
\be 2\,p\,{\cal H} =\left( \begin{array}{cccccccc} m_{D}^{2}(n+1)
- 2p(1 -\Omega) V_m & m_{D}/R & - (m_{D}/R)&
2 (m_{D}/R) & - 2 (m_{D}/R) & \cdots & n (m_D/R) & - n (m_D/R) \\[0.5em]
(m_{D}/R)  & 1/R^2 & 2pV_{G}^{(2)} & 2pV_{G}^{(1)} &
2pV_{G}^{(3)}  &
\cdots &  2pV_G^{(n-1)}  & 2pV_G^{(n+1)}\\[0.5em]
 - (m_{D}/R) &  2pV_{G}^{(2)} & 1/R^2  &  2pV_{G}^{(3)} &
2pV_{G}^{(1)} & \cdots & \cdots & \cdots   \\[0.5em]
2 (m_{D}/R)  & 2pV_{G}^{(1)}  & 2pV_{G}^{(3)} &  4/R^2 &
2pV_{G}^{(4)}
& \cdots &\cdots & \cdots \\[0.5em]
- 2 (m_{D}/R) & 2pV_{G}^{(3)}  & 2pV_{G}^{(1)} &  2pV_{G}^{(4)} &
4/R^2 &
\cdots & \cdots & \cdots \\[0.5em]
\cdots &\cdots &\cdots & \cdots  & \cdots &\cdots &\cdots & \cdots \\
 n (m_D /R) &  2pV_G^{(n-1)} & \cdots & \cdots &\cdots  & \cdots &
n^2/R^2 &  2pV_G^{(2n)}\\[0.5em]
 - n (m_D /R) &  2pV_G^{(n+1)} & \cdots &\cdots &\cdots &\cdots
&  2pV_G^{(2n)} & n^2/R^2
\end{array} \right) \label{M} \, .
\ee \vspace{0.5cm}
\end{figure*}

In \eq{ev} we left out
 common (kinetic and potential) terms that appear in the diagonal of the matrix~\eq{M},  and contributions
quadratic in the potentials.
In \eq{M},
\be
V_G^{(n)} \equiv \frac{1}{2 R} \int^{+R}_{-R} \cos\,
 (\pi n y/R)\, V_G(y)\, \di y
\ee
is the $n$-th Fourier component of the gravitational potential energy.

To gauge the magnitude of this new gravitational term, we have written
in \eq{M} the difference between the potential in $y=0$ and its
zero mode as
\be
V_{\rm G} (0) - V_{\rm G}^{(0)} \equiv - \Omega \; V_{\rm m} \, ,
\ee
with
\be
\Omega \simeq   10^5\; \left( \frac{\rm 1\, TeV}{M_5} \right)^3
\left( \frac{\rho}{\rm 100\, \mu m} \right)
\left( \frac{p}{\rm 1\, MeV} \right) \frac{\xi_{N}}{\xi_e}
\, . \label{omega}
\ee

For a situation in which $V_m=V_G^{(n)}=0$ in \eq{M}, the oscillation
in vacuum between the standard-model neutrino
and the $n$-th state $N^{(n)}$ of the tower of KK sterile neutrinos, with
masses given by $n/R$, is governed by the mixing angle
\be
\tan 2\, \theta^{(n)}_0 = \frac {2\, m_D \, (n/R)}{(n/R)^2 - m_D^2 (n+1)} \, .
\ee

Consider now the effect of matter, but neglecting for a moment gravitional
interactions.
As neutrinos enter space filled by matter with constant density,
the mixing angle becomes
\be\label{weakan}
\tan 2\, \theta^{(n)}_m =
\frac {\sin 2\, \theta^{(n)}_0} {\cos 2\, \theta^{(n)}_0
- 2\,p \,V_m\, (R/n)^2} \,.
\ee
The weak potential modifies the mixing angle, leading to a
possible resonance,  in a way that strongly resembles the
usual MSW effect.

Let us finally switch on gravity in our system.
Neglecting for the moment  off-diagonal terms in \eq{M} originating from
higher Fourier modes of the gravitational potential, we can write the mixing
angle as
\be
\tan 2\, \theta^{(n)}_{\small{(m+G)}} =
\frac {\sin 2\, \theta^{(n)}_0}
{\cos 2\, \theta^{(n)}_0 -2\, (1-\Omega)\,p\,V_m\, (R/n)^2} \label{full}
\, .
\ee
If the sterile neutrinos were constrained on the four dimensional
world, $\Omega$, as well as  all
the coefficients $V_G^{(n)}$, would vanish and there would be no
effect.
If $\Omega \gg 1$,  the effective mixing angle in
matter of $\nu_L$ and, for instance, the first KK mode $N^{(1)}$ is
suppressed by the factor $\Omega^{-1}$:
\be
\tan 2\, \theta^{(1)}_{(m+G)} \sim \Omega^{-1} \tan 2\,
\theta^{(1)}_m \, . \label{last}
\ee
Accordingly, gravity could decouple
the two neutrino modes.
Let us stress that the sign of the gravitational
potential is opposite to the weak potential in \eq{m}:
when $V_G$ is large enough to flip the sign of matter effects,
it is no more possible to satisfy a resonance condition,
since we consider  small mixing angles $\theta_{0}$.

This fact allows also for the possibility of a compensation, for
a choice  of  parameters that leads to $\Omega \simeq 1$,
 between weak and gravitational contributions.
In this case, even in presence of matter, the effective mixing angles
result the same as in  vacuum.

Let us stress an important difference between gravitational and
weak interactions: while the former are the same for particles and
antiparticles, the presence of a medium makes the latter act
differently between neutrinos and antineutrinos. In this sense,
if gravitational effects are relevant, they can be factorized in
a comparative study between neutrino and antineutrino fluxes, and
it is possible to extract indepent information on weak and
gravitational effects.

The same results hold by using the full matrix (\ref{M}), off-diagonal
Fourier terms included. We have checked
numerically that the diagonalization of the matrix, for the
values of $\Omega$ discussed above, gives rise to
eigenvectors in which $\nu_L$ is, for all practical
purposes, decoupled from the bulk-neutrino modes.

\section{Model dependent considerations}\label{applications}

Having established that gravity does affect neutrino oscillations in models where the sterile neutrinos propagate through the extra dimensions, let us now discuss the possible relevance of this effect.

 Our discussion can be seen from two complementary points of view:
either as the search for models in which  gravitational effects are
important or as the definition of models which are safe from such effects. Indeed, even
showing that gravity---although present--- does not significantly modify the physics is relevant, since
most neutrino physics is valuable exactly because
neutrinos only interact via electroweak forces.

The  first example  we consider is that of solar neutrinos in the  $D=5$ model with one
  extra-dimension compactified on a circle of radius $R$~\cite{extra} that we used as a template in the previous sections.
 In this case, we can use directly  the computation performed in
Section~(\ref{matter}), identifying $\rho \simeq R$.
The four dimensional effective Planck scale $M_{\rm Pl}$ is connected
with the foudamental scale $M_{5}$ by
\be
M^2_{\rm Pl} = 2\, R\, M_{5}^3 \label{c} \, .
\ee
The largest allowed radius is at the sub-millimiter scale, so that
$M_{5} \sim 10^5$ TeV.

 Considering
typical energies for neutrinos  $p \sim$ few MeV, and an averaged
value for the solar density, we obtain for $\Omega$ the value $10^{-10}$.
This means that for all practical purposes, in \eq{M} one can
set to zero $V^{n}_{G}$ and $\Omega$: gravity is negligible in respect to the weak interaction effects, even in presence of
sterile KK states.


How about the more general case of models based on more complicated space-times?
One should repeat  the computation of
Section~\ref{matter}, substituting $G_{5}$ in~\eq{int}  with
the new coupling, and the appropriated graviton two-point
correlator in~\eq{pot}.

For two flat extra dimensions of the same size, the two point correlator
goes like $r^{-3}$ instead of $r^{-2}$. This implies that
 the potential in \eq{pot} yields
a logarithmic dependence on $R$ which makes the effect smaller. Further
additional dimensions make the potential even weaker. In this case,
even phenomena involving ultra high-energy neutrinos are not affected.

Another case in which the two point correlator is known is the
Randall-Sundrum model~\cite{RS}: the power law dependence of the
gravitational potentials are similar to the one of $D=6$ ADD model,
and the same considerations hold, even if the fundamental coupling
of the model is lowered with respect to the original idea of
RS, in which Planck scale gravity inhabits the bulk.

We therefore conclude that, at least for these models for which
we know how to compute the gravitational potential, we do not
expect gravity to play a role at the energies of solar neutrinos,
because of the large value of the scale $M_5$ suppresses their
gravitational interactions.

 However,
even for  $M_{5}$ of the order of $10^{5}$ TeV,
there could be important effects for neutrino physics in other situations.
Consider, for instance, ultra-high energy
neutrinos produced by astrophysical objects like gamma ray bursters
or active galactic nuclei: their oscillations,
also  to sterile neutrinos,
in presence of matter effects have been discussed~\cite{LS}.
If any sterile neutrino lives in the bulk, a Dirac equation similar to
the one discussed for the solar neutrinos applies, in which matter effects
are due to the interstallar medium.  The suppression in
\eq{omega}  coming from the large value of $M_5$ is compensated, in
this case, by the energy $p$ which can be as large as $10^9$ TeV,
even if the densities are much smaller: the
gravitational terms effectively compete with weak interactions in
determining matter effects in
the oscillations.

 Possible gravitational interactions of ultra-high
energy  neutrinos has also been discussed for the physics
of scattering processes in~\cite{NS}.

\section{Model independent considerations}
\label{modind}

The actual value of $M_{5}$ in a generic model
should be fixed by the dynamics of the
system which gives rise to the compactification geometry
and by the implied relation between the effective
extra dimensional couplings and $G_N$.
This is an open problem in extra-dimensional models,
and, in general, $M_{5}$ is neither given by \eq{c} nor known,
apart from few peculiar examples, as the RS model.

For this reason it is also useful to follow a purely
phenomenological approach in which $\rho$ and $M_5$ are only
restricted by the experimental bounds. We assume
un-su\-ppres\-sed gravity to be localized
at very short distances, smaller than $\rho$,
while for $r \gg \rho$ the usual Newtonian
regime (up to subleading long-range corrections, due to the
exchange of heavy KK states) is recovered.

In general, from the combined
gravimetrics~\cite{exp} and particle-physics~\cite{lep}
experimental bounds we can extract
the scale $M_5$ and the effective range $\rho$ of gravity which we
use as the cut off in \eq{pot}. Since these bounds are crucial,
 we
summarize and discuss them in the following subsections.

\subsection{Searches for non-newtonian gravity}

While Newtonian gravity above the centimeter is well confirmed~\cite{AdR},
its short distance behavior is still under active scrutiny.
All experiments, regardless of the actual apparatus,
set a bound on non-Newtonian interactions from the absence of
deviations between the force measured at distance $r^{\ast}$ and
the predicted one.

The bound is usually given in terms of the parameters $\alpha$ and $\lambda$
according to the two-body potential
\be
\left. V(r)\right|_{r^*}= \left. \frac{G_N m_1 m_2}{r} \Biggl[
1+\alpha_{G}\, e^{-r/\lambda}  \label{pot_exp}
\Biggr]\right|_{r^*} \, ,
\ee
where $G_N$ is the Newton constant in four space-time dimensions.
Because of  the exponential behavior,
the best sensitivity is achieved in the range $\lambda \sim r^{\ast}$.
Currently,
experiments testing Van der Waals forces  are sensitive to the range
$r^{\ast} \sim 1.5 \div 130$ nm~\cite{vdW};
Casimir-force experiments explore
$r^{\ast} \sim 0.02 \div 6$ $\mu$m , $r^{\ast}$ being here the
distance between dielectrics or metal
surfaces~\cite{casimir1,casimir2} or  up  to mm
by means of a torsion pendulum\cite{casimir3}.
Cavendish-type experiments, in which the gravitation force is directly
measured, are sensitive to $r^{\ast} > 1$ mm~\cite{torsion}.

The exclusion regions
thus determined are convex curves around the
distance $r^*$ at which the experiment is
performed. The sensitivity of the experiments rapidly
decreases at smaller wave-lenghts.
The combined exclusion regions obtained by these searches, for the relevant
distances, are shown as grey areas delimited by black curves in
Fig.~\ref{fig1}.
\vskip1.8em
\subsection{Gravitational potential in models with large extra dimensions.}

The two-body potential in models with $\delta$ flat extra dimensions can be
parameterized (for $r$ less than $R^*$,
the characteristic compactification length) as
\be
V_\delta (r)=\frac{G_N m_1 m_2}{r}
\left( \frac{a_\delta}{r}
\right)^{\delta} \label{pot_s} \, .
\ee
In \eq{pot_s}
\be
a_\delta =
(G^{(\delta)}/G_N)^{1/\delta}= \frac{2\,\pi}{M_{f}}\;
\left(\frac{4\,\pi}{\Omega_{\delta}}
\frac{M_{\rm P}^2}{M_{f}^{2}} \right)^{1/ \delta} \, , \label{adelta}
\ee
where we define
$M_{\rm P} \equiv 1/\sqrt{G_N} = 1.22 \times 10^{16}$ TeV. In \eq{adelta},
$\Omega_{\delta}=2 \pi^{(3+\delta)/2}/ \Gamma[(3+\delta)/2]$
and $M_f$ is the scale of the
effective theory.
For distances larger than $R^*$,
the potential in \eq{pot_s} is replaced by the usual
Newtonian potential plus exponentially small corrections:
\be
V_\delta (r) =
\frac{G_N m_1 m_2}{r} \Biggl[
1+\alpha_\delta\, e^{-r/R^*} +  \cdots  \label{pot_l}
\Biggr]  \, .
\ee
In \eq{pot_l},
the value of $\alpha_\delta$ depends on the
compactification choice and is of the order of the number of
extra dimensions~\cite{expo}.

It is important to bear in mind that
\eq{pot_l} depends on the way the extra dimensions are
treated in the process of compactification
while \eq{pot_s} only relies on Gauss law and is therefore
compactification independent.

When the experimental bounds parameterized by \eq{pot_exp}
are plotted (on a logarithmic scale in the
$(\alpha-\lambda)$-plane)
against \eq{pot_l}, a single point is obtained
at $\alpha_G = \alpha_\delta$ and
$\lambda = R^{\ast}$. For instance,
models with extra-dimension compactified on
torii with equal radii $R^{\ast}$
predict $\alpha_{\delta}=2\delta$.
This leads to the known bound $R^{\ast}\gtap$ few $10^{-4}$ m,
as can be deduced by Fig.~\ref{fig1}, with $R^{\ast}=\lambda$.


\vskip1.5em
\subsection{Compactification-independent  bounds from particle physics.}
Contrarily to short-distance gravity measurements,
particle-physics measurements 
only constrain the effective gravitational coupling $G^{(\delta)}$
by means of the bound on $M_f$.
The independence from $r$ is manifest in the $(4+\delta)$-dimensional
 theory, which probes distances
much smaller than the compactification radius $R^*$, and recovered in the
 4-dimensional computation  after resumming over  the Kaluza-Klein states.

For this reason, the relationship
obtained by comparing \eq{pot_exp} and \eq{pot_s},
must be valid for any choice of $r$ (as long as $r \ltap R^*$)
and gives the stringiest bound at the minimum.
 Therefore, the curve of exclusion is
found to be
\be
 \alpha_G (\lambda) \le \mathop{\mathrm{min}}_{r} \Big\{ \Big[ \left(\frac{a_{\delta}}{r} \right)^{\delta}-1 \Big]e^{r/\lambda} \Big\} \, .
 \label{5}
\ee
To find the curve given by \eq{5}, we must solve the polynomial
equations
obtained by the minimalization procedure.
Exact solutions exist for $\delta < 4$; however,
 for all practical purposes, approximated solutions can be found by elementary
calculus for any $\delta$. The exclusion region is given by the lines
\be
\alpha_{G}(\lambda) = \left\{ \begin{array}{ll}
\left[ \left( a_{\delta}/\delta \lambda \right)^\delta - 1 \right] e^\delta &
\textrm{for $\lambda < \lambda_{max}$} \\ & \\
 \alpha_{min} \equiv \delta \, e^{\delta+1} &
\textrm{for $\lambda \geq \lambda_{max}$} \, ,
\end{array} \right. \label{6}
\ee
where $\lambda_{max} = (1+\delta)^{-(1+\delta)/\delta}a_{\delta}$. The
value $\lambda_{max}$ is reached when no real solution can be
found. The exclusion region is extended for $\lambda > \lambda_{max}$
by taking   smaller values of $M_f$ (already excluded)  for which the solution is
translated to larger values of $\lambda$ while still ending at the same
(constant) value of $\alpha_{min}$.

In collider physics,
the most effective channel at both LEP and Tevatron is that in which
virtual gravitons take part in dilepton or diphoton production.
Production of real graviton gives less stringent bounds. Whenever the
bound depends on the sign of the potential we have taken the lesser bound.
Recent reviews of all these
bounds can be found in Ref.~\cite{rev}.

We have summarized in
Table~\ref{hep-bounds} the best bounds from particle physics.
\begin{center}
\begin{table}[ht]
\caption{Particle physics bounds on $M_f$. The numbers reported are
the constrains in TeV for the first few large extra dimensions
$\delta$. Missing entries were not reported in the literature.}
\label{hep-bounds}
\begin{center}
\begin{tabular}{c|ccc|c}
\hline
 & \multicolumn{3}{c|}{$\delta$} &  \\
                   \cline{2-5}
measurement             & 1  & 2   & 3      & reference  \\
                    \cline{2-5}
\hline
LEP          &1.2& 1.2 & 1.2 &   \cite{l3} \\
Tevatron I   &- &1.5  & 1.5 &   \cite{d0,tevatron} \\
\hline
Tevatron IIb &- &3.5  & 3.0 &   \cite{tevatron} \\
LHC          &- & 13 & 12 &  \cite{tevatron} \\
\hline
\end{tabular}
\end{center}
\end{table}
\end{center}
While precision measurements and
collider bounds from production of real gravitons depend on the
number of extra dimensions, those from virtual graviton processes at
colliders are (almost,
for certain parameterizations) independent. Bounds from oblique parameters are
potentially very restrictive but are plagued by infrared divergences
which make the final result rather uncertain~\cite{rattazzi}. For this reason we will
not use them.

Even though a degree
of uncertainty remains in these calculations because of the cut-off
dependence (and because of different parameterizations),
the bounds work on order of magnitudes and are therefore sufficiently
reliable as they stand. In particular, we neglect small
discrepancies between different approaches~\cite{colliders}.

We keep in Table~\ref{hep-bounds} also the $\delta=$1
 case, even though it is often considered ruled out.
This is true only
after having assumed a specific compactification geometry
and we want
to use the particle-physics constraints irrespectively
of this additional assumption.

%

\begin{figure*}
\includegraphics{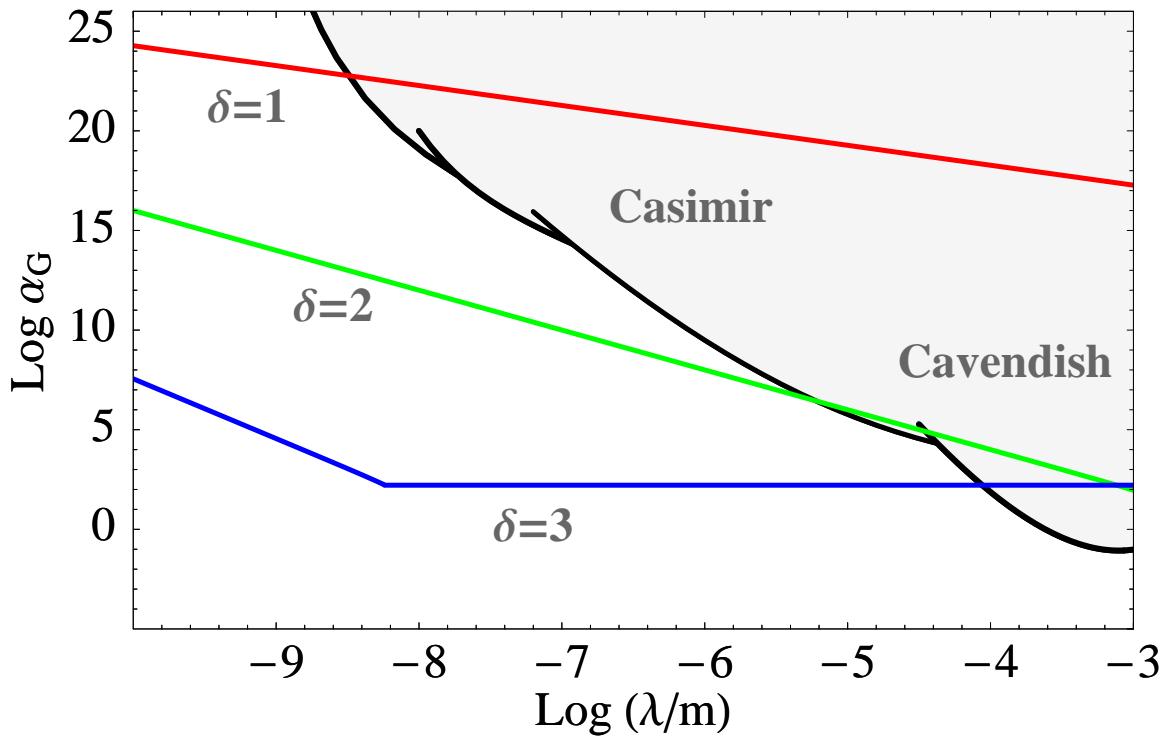}
\caption{Bounds $\alpha_G$ vs.\ $\lambda$ for $\delta =$ 1, 2 and 3 from
current particle-physics tests (see Table~\ref{hep-bounds}).
The thicker curved lines are the best available bound from
non-Newtonian gravity experiments and the grey area is the range
excluded by them. }
\label{fig1}
\end{figure*}

Given the particle-physics bounds in Table~\ref{hep-bounds} and \eq{6},
we obtain the curves in Fig.~\ref{fig1}, where
the respective exclusion regions (the
area above the lines) are presented for the
first few extra dimensions.

In using these bounds,
the values for $\alpha_G$ and
$\lambda$ of a specific model
must be plotted against the bounds of the corresponding effective
theory at $r \sim \lambda$,
the space dimension of which is not necessarily that of the
fundamental theory.

Figure~\ref{fig1}  shows that for $\delta > 2$
particle-physics bounds, in particular
those coming from collider physics, are
various orders of magnitude stronger than direct searches for
non-Newtonian gravity below the mm.
In other words, if any deviation is ever found
in these experiments, it
will not be possible to explain it in terms of
large extra-dimension models.

On the contrary, for $\delta=1$ in the range $\lambda \gtap 1$ nm
Casimir and Cavendish-like experiments are the most sensitive
and rule out a large amount of parameter space, while particle physics
is relevant only at much shorter distances. Notice that the bounds
still allow a strong gravity coupling (of the order of $1/({\rm TeV})^3$)
up to few nm as long as it
then decreases fast enough to match
the long distance regime, in order to
satisfy bounds from gravimetric experiments.

The conclusion of this analysis of the experimental bounds is that,
for most of the range of parameters we are interested in,
high-energy experiments give the strongest bounds on the size
of space-time extra dimensions and gravity strength. Coming
back now to the impact of gravity on neutrino physics, even
by taking into account these bounds, and therefore by
considering distances shorter than
$\rho \sim 10^{-3}\mu$m and taking the
limit of $M_5 \geq 1.2$ TeV, the crucial factor
$\Omega$ in (\ref{omega})  can be as large as ten
for neutrino energies in the range of solar neutrino physics.

\section{Conclusions}
\label{conclusions}

We have shown that whenever fields with the same quantum
numbers are allowed to propagate in different portions
of space-time, the four dimensional theory exhibits
 an effective violation of the principle of equivalence.
This implies that, in principle, gravity could play a role in
flavor violating phenomena such as neutrino oscillations
in matter, due to the non-universality
of the gravitational coupling.

Furthermore, large extra dimension scenarios could imply a strong
enhancement of the coupling at short distances, where the
space-time is effectively higher dimensional. We focused on the
case of just one extra dimension, and discussed explicitly the
conditions under which such an effect is relevant for neutrino
physics, even though our discussion is only at the level of
orders of magnitude and matter is taken to have a constant
density distribution.

Although the model we considered is not compatible with recent
experimental results, its simplicity allowed us to show how to
estimate quantitatively possible effects of gravitational
interactions in neutrino experiments.

Accordingly, the presence of the  gravitational potential may, in
principle, drastically change the values of the parameters used in
fitting the experimental data and the interpretation of a
resonance solution; moreover, it can produce a peculiar
distortion of the neutrino energy spectrum because of the extra
energy dependence.

In the simplest case of  compactification on a flat manifold,
the effect of gravity is many orders of magnitude
too weak for observation, and only weak interactions play
an important role for solar neutrino oscillation. Still,
gravity could be relevant for more exotic phenomena,
for example in the study of ultra high energy neutrinos in cosmic rays,
where the effect is enhanced by the peculiar energy
dependence of gravitational interactions. In this case, the generic
effect of gravitational interactions  is to
 suppress the effective mixing angles in matter.

To conclude, let us notice that gravity could also be relevant
for sterile neutrino physics at lower energies if the mechanism
of compactification is more subtle than that on torii, and
provided all the bounds on sterile neutrino physics are satisfied.

\section*{Acknowledgments}

MP and GT  would like to thank  G. Dvali, C. Lunardini,
S. Pascoli and A.~Yu.\ Smirnov for useful discussions. This work was
partially supported by the European TMR Networks
HPRN-CT-2000-00148 and 00152.

\vspace*{0.6cm}

%
%
%
%
%

\end{document}